\documentclass[11pt,a4paper,english]{article}
\pdfoutput=1

\usepackage{lmodern}
\renewcommand{\sfdefault}{lmss}

\usepackage{color}
\usepackage{float}
\usepackage{mathtools}
\usepackage{amsmath}
\usepackage{amsthm}
\usepackage{amssymb}
\usepackage{graphicx}

\makeatletter

\newif\ifprstyle

\ifx\affiliation\undefined
  \prstylefalse
\else
  \prstyletrue
  \newif\ifnotoc
  \notoctrue
\fi

\ifprstyle
  \usepackage[
    colorlinks=true, pdfa=true,
    urlcolor=blue,   anchorcolor=blue,
    citecolor=blue,  filecolor=blue,
    linkcolor=blue,  menucolor=blue,
    pagecolor=blue,  linktocpage=true
  ]{hyperref}
\else
  \usepackage{jheppub}
  \newcommand{\email}[1]{\emailAdd{#1}}
\fi

\ifx\ifCrop\undefined

  \divide\@tempdima\baselineskip
  \@tempcnta=\@tempdima
\else

  \ifx\ifPresentation\undefined
     \usepackage[paperwidth=171.2mm,paperheight=261.2mm]{geometry}
  \else
     \usepackage[paperwidth=330mm,paperheight=233mm]{geometry}
  \fi

  \divide\@tempdima\baselineskip
  \@tempcnta=\@tempdima
  \ifx\ifPresentation\undefined
    \hypersetup{pdfstartview={Fit},pdfpagelayout={TwoPageRight}}
  \else
    \hypersetup{pdfstartview={Fit},pdfpagelayout={SinglePage}}

    \usepackage{everypage}
    \AddEverypageHook{\tikz[remember picture,overlay]{%
      \draw [gray!10] ($(current page.east)+(-130mm,+180mm)$)
         -- ($(current page.east)+(-130mm,-100mm)$);}}
  \fi
\fi

\hypersetup{
    urlcolor=black!40!blue,   anchorcolor=black!40!blue,
    citecolor=black!40!blue,  filecolor=black!40!blue,
    linkcolor=black!40!blue,  menucolor=black!40!blue,
    pagecolor=black!40!blue
}

\usepackage{bookmark}

\usepackage{amsmath,amsfonts,amssymb,bm,cancel}

\usepackage{color}
\usepackage{colortbl}

\ifprstyle
  
\else
  
\fi

\newcommand{\bSe}{\begin{subequations}}
\newcommand{\eSe}{\end{subequations}}

\newcommand{\bWe}{\begin{widetext}}
\newcommand{\eWe}{\end{widetext}}

\allowdisplaybreaks

\usepackage{tikz}
\usetikzlibrary{shapes,arrows}
\usetikzlibrary{calc}
\usetikzlibrary{positioning}
\usetikzlibrary{decorations.pathreplacing}
\usetikzlibrary{shapes.misc}

\usepackage[scr,scaled=1.1]{rsfso}
\usepackage{eucal}

\DeclareMathAlphabet{\mathsfit}{\encodingdefault}{\sfdefault}{m}{sl}
\SetMathAlphabet{\mathsfit}{bold}{\encodingdefault}{\sfdefault}{bx}{sl}

\renewcommand\floatc@plain[2]{\setbox\@tempboxa\hbox{{\@fs@cfont #1.} #2}%
\ifdim\wd\@tempboxa>\hsize {\@fs@cfont #1.} #2\par
\else\hbox to\hsize{\hfil\box\@tempboxa\hfil}\fi}

\ifprstyle\else
\fi

\usepackage[framemethod=tikz]{mdframed}

\mdfdefinestyle{title}{%
innerleftmargin=10,innerrightmargin=5,
leftmargin=25mm,skipabove=\topsep,skipbelow=\topsep,
leftline=false,topline=false,rightline=false,bottomline=false,outerlinewidth=0.1pt,
backgroundcolor=gray!0,linecolor=gray!20
}

\newcommand{\emitFrontMatter}{
\makeatletter
\compress
\renewcommand\afterLogoSpace{}
\renewcommand\afterSubheaderSpace{}
\renewcommand\afterProceedingsSpace{}
\toccontinuoustrue
\renewcommand\ps@titlepage{}
\pagestyle{empty}
\thispagestyle{titlepage}
\setcounter{page}{0}
{\LARGE\flushleft\sffamily\bfseries\@title\par}
\begin{mdframed}[style=title]%
{\bfseries\raggedright\sffamily\the\auth@toks\par}
\afterAuthorSpace
\ifaffil\begin{list}{}{%
\setlength{\leftmargin}{3mm}%
\setlength{\labelsep}{0pt}%
\setlength{\itemsep}{\affiliationsSep}%
\setlength{\topsep}{-\parskip}}
\itshape\small%
\the\affil@toks
\end{list}\fi   
\noindent\hspace{3mm}\begin{minipage}[l]{.89\textwidth-3mm}%
\begin{flushleft}
\textit{E-mail:} \the\email@toks
\end{flushleft}
\end{minipage}%
\par\medskip
\noindent\bgroup\renewcommand{\baselinestretch}{0.95}\selectfont%
\textbf{Abstract.}\ \myAbstract
\egroup
\end{mdframed}%
\setcounter{footnote}{0}
\pagestyle{myplain}\pagenumbering{arabic}
\makeatother
\hrule
\bigskip
}

\usepackage{titlesec}

\makeatother

\usepackage{babel}
\begin{document}

\newif\ifColors

\newif\ifShowSigns 

\ifx \ii \undefined

\definecolor{red}{rgb}{1,0,0.1}
\definecolor{green}{rgb}{0.0,0.6,0}
\definecolor{blue}{rgb}{0.1,0.1,1}
\definecolor{brown}{rgb}{0.6,0.3,0}
\definecolor{orange}{rgb}{0.8,0.3,0}
\definecolor{magenta}{rgb}{0.9,0.1,1}

\newcommand\nPlusOne{$N$+1}

\renewcommand\tilde[1]{\mkern1mu\widetilde{\mkern-1mu#1}}

\global\long\def\ii{\mathrm{i}}
\global\long\def\ee{\mathrm{e}}
\global\long\def\dd{\mathrm{d}}
\global\long\def\ppi{\mathrm{\pi}}
\global\long\def\tr{\mathsf{{\scriptscriptstyle T}}}
\global\long\def\Tr{\operatorname{Tr}}
\global\long\def\op#1{\operatorname{#1}}
\global\long\def\dim{\operatorname{dim}}
\global\long\def\diag{\operatorname{diag}}
\global\long\def\Lie{\mathrm{\mathscr{L}}}

\global\long\def\mfrac#1#2{\frac{\raisebox{-0.45ex}{\scalebox{0.9}{#1}}}{\raisebox{0.4ex}{\scalebox{0.9}{#2}}}}
\global\long\def\mbinom#1#2{\Big(\begin{array}{c}
 #1\\[-0.75ex]
 #2 
\end{array}\Big)}

\global\long\def\tud#1#2#3{#1{}^{#2}{}_{#3}}
\global\long\def\tdu#1#2#3{#1{}_{#2}{}^{#3}}

\global\long\def\qvf{\xi}
\global\long\def\ixA{a}
\global\long\def\ixB{b}
\global\long\def\ccVar{\mathcal{C}}

\global\long\def\lidx#1{\ ^{(#1)}\!}

\global\long\def\gSector#1{{\color{black}#1}}
\global\long\def\fSector#1{{\color{black}#1}}
\global\long\def\hSector#1{{\color{black}#1}}
\global\long\def\sSector#1{{\color{black}#1}}
\global\long\def\lSector#1{{\color{black}#1}}
\global\long\def\mSector#1{{\color{black}#1}}
\global\long\def\hrColor#1{{\color{black}#1}}
\global\long\def\VColor#1{{\color{black}#1}}
\global\long\def\KColor#1{{\color{black}#1}}
\global\long\def\KVColor#1{{\color{black}#1}}
\global\long\def\pfSector#1{{\color{black}#1}}

\global\long\def\gMet{\gSector g}
\global\long\def\gSp{\gSector{\gamma}}
\global\long\def\gK{\gSector K}
\global\long\def\gE{\gSector e}
\global\long\def\gD{\gSector D}
\global\long\def\gR{\gSector R}
\global\long\def\gCS{\gSector{\Gamma}}
\global\long\def\gVse{\gSector{V_{g}}}
\global\long\def\gTse{\gSector{T_{g}}}
\global\long\def\gEinst{\gSector{G_{g}}}
\global\long\def\gRicci{\gSector{R_{g}}}
\global\long\def\gCC{\gSector{\mathcal{C}}}
\global\long\def\gCE{\gSector{\mathcal{E}}}
\global\long\def\gCD{\gSector{\nabla}}
\global\long\def\gPi{\gSector{\pi}}

\global\long\def\fMet{\fSector f}
\global\long\def\fSp{\fSector{\varphi}}
\global\long\def\fK{\fSector{\tilde{K}}}
\global\long\def\fE{\fSector m}
\global\long\def\fD{\fSector{\tilde{D}}}
\global\long\def\fR{\fSector{\tilde{R}}}
\global\long\def\fCS{\fSector{\tilde{\Gamma}}}
\global\long\def\fVse{\fSector{V_{f}}}
\global\long\def\fTse{\fSector{T_{f}}}
\global\long\def\fEinst{\fSector{G_{f}}}
\global\long\def\fRicci{\fSector{R_{f}}}
\global\long\def\fCC{\fSector{\widetilde{\mathcal{C}}}}
\global\long\def\fCE{\fSector{\widetilde{\mathcal{E}}}}
\global\long\def\fCD{\fSector{\widetilde{\nabla}}}
\global\long\def\fPi{\fSector p}
\global\long\def\fPi{\fSector{\tilde{\pi}}}

\global\long\def\gLapse{\gSector{\alpha}}
\global\long\def\gShift{\gSector{\beta}}
\global\long\def\gShiftVec{\gSector{\beta}}

\global\long\def\fLapse{\fSector{\tilde{\alpha}}}
\global\long\def\fShift{\fSector{\tilde{\beta}}}
\global\long\def\fShiftVec{\fSector{\tilde{\beta}}}

\global\long\def\gKappa{\gSector{\kappa_{g}}}
\global\long\def\gKappainv{\gSector{\kappa_{g}^{-1}}}
\global\long\def\Mg{\gSector{M_{g}^{d-2}}}

\global\long\def\fKappa{\fSector{\kappa_{f}}}
\global\long\def\fKappainv{\fSector{\kappa_{f}^{-1}}}
\global\long\def\Mf{\fSector{M_{f}^{d-2}}}

\global\long\def\grho{\gSector{\rho}}
\global\long\def\gjota{\gSector j}
\global\long\def\gJota{\gSector J}

\global\long\def\frho{\fSector{\tilde{\rho}}}
\global\long\def\fjota{\fSector{\tilde{j}}}
\global\long\def\fJota{\fSector{\tilde{J}}}

\global\long\def\grhom{\gSector{\rho^{\mathrm{m}}}}
\global\long\def\gjotam{\gSector{j^{\mathrm{m}}}}
\global\long\def\gJotam#1{\gSector{J_{#1}^{\mathrm{m}}}}

\global\long\def\frhom{\fSector{\tilde{\rho}^{\mathrm{m}}}}
\global\long\def\fjotam{\fSector{\tilde{j}^{\mathrm{m}}}}
\global\long\def\fJotam#1{\fSector{\tilde{J}_{#1}^{\mathrm{m}}}}

\global\long\def\grhob{\gSector{\rho^{\mathrm{b}}}}
 \global\long\def\gjotab{\gSector{j^{\mathrm{b}}}}
 \global\long\def\gJotab{\gSector{J^{\mathrm{b}}}}

\global\long\def\frhob{\fSector{\tilde{\rho}^{\mathrm{b}}}}
 \global\long\def\fjotab{\fSector{\tilde{j}^{\mathrm{b}}}}
 \global\long\def\fJotab{\fSector{\tilde{J}^{\mathrm{b}}}}

\global\long\def\grhoeff{\gSector{\rho_{\mathrm{eff}}}}
 \global\long\def\gjotaeff{\gSector{j_{\mathrm{eff}}}}
 \global\long\def\gJotaeff{\gSector{J_{\mathrm{eff}}}}

\global\long\def\frhoeff{\fSector{\tilde{\rho}_{\mathrm{eff}}}}
 \global\long\def\fjotaeff{\fSector{\tilde{j}_{\mathrm{eff}}}}
 \global\long\def\fJotaeff{\fSector{\tilde{J}_{\mathrm{eff}}}}

\global\long\def\gAlpha{\gSector{\alpha}}
\global\long\def\gBeta{\gSector{\beta}}
\global\long\def\gEA{\gSector A}
\global\long\def\gEB{\gSector B}
\global\long\def\fAlpha{\fSector{\tilde{\alpha}}}
\global\long\def\fBeta{\fSector{\tilde{\beta}}}
\global\long\def\fEA{\fSector{\tilde{A}}}
\global\long\def\fEB{\fSector{\tilde{B}}}

\global\long\def\sEtau{\mSector{\tau}}
\global\long\def\sESigma{\mSector{\Sigma}}
\global\long\def\sER{\mSector R}

\global\long\def\cphi{\gSector{\psi}}
\global\long\def\gcphi{\gSector{\psi}_{\gMet}}
\global\long\def\fcphi{\fSector{\psi}_{\fMet}}

\global\long\def\gblab{\gSector{\mathrm{b}}}
\global\long\def\gmlab{\gSector{\mathrm{m}}}
\global\long\def\fblab{\fSector{\mathrm{b}}}
\global\long\def\fmlab{\fSector{\mathrm{m}}}

\global\long\def\pfrho{\pfSector{\rho}_{\pfSector 0}}
\global\long\def\pfD{\pfSector{\hat{D}}}
\global\long\def\pfS{\pfSector{\hat{S}}}
\global\long\def\pftau{\pfSector{\hat{\tau}}}
\global\long\def\pfu{\pfSector u}
\global\long\def\pfv{\pfSector{\hat{v}}}
\global\long\def\pfW{\pfSector w}
\global\long\def\pfh{\pfSector h}
\global\long\def\pfeps{\pfSector{\epsilon}}
\global\long\def\pfP{\pfSector P}

\global\long\def\Proj{\operatorname{\perp}}
\global\long\def\gProj{\gSector{\operatorname{\perp}_{g}}}
\global\long\def\fProj{\fSector{\operatorname{\perp}_{f}}}
\global\long\def\hProj{\hSector{\operatorname{\perp}}}
\global\long\def\prho{\boldsymbol{\rho}}
\global\long\def\pjota{\boldsymbol{j}}
\global\long\def\pJota{\boldsymbol{J}}

\global\long\def\sgn{\gSector{\mathsfit{n}{\mkern1mu}}}
\global\long\def\sgD{\gSector{\mathcal{D}}}
\global\long\def\sgQ{\gSector{\mathcal{Q}}}
\global\long\def\sgV{\gSector{\mathcal{V}}}
\global\long\def\sgU{\gSector{\mathcal{U}}}
\global\long\def\sgB{\gSector{\mathcal{B}}}

\global\long\def\sfn{\fSector{\tilde{\mathsfit{n}}{\mkern1mu}}}
\global\long\def\sfD{\fSector{\widetilde{\mathcal{D}}}}
\global\long\def\sfQ{\fSector{\widetilde{\mathcal{Q}}}}
\global\long\def\sfV{\fSector{\widetilde{\mathcal{V}}}}
\global\long\def\sfU{\fSector{\widetilde{\mathcal{U}}}}
\global\long\def\sfB{\fSector{\widetilde{\mathcal{B}}}}

\global\long\def\sgW{\gSector{\mathcal{W}}}
\global\long\def\sgQU{\gSector{(\mathcal{Q\fSector{{\scriptstyle \widetilde{U}}}})}}

\global\long\def\sfW{\fSector{\tilde{\mathcal{W}}}}
\global\long\def\sfQU{\fSector{(\mathcal{\widetilde{Q}\gSector{{\scriptstyle U}}})}}

\global\long\def\hMet{\hSector h}
\global\long\def\hSp{\hSector{\chi}}
\global\long\def\hLapse{\hSector H}
\global\long\def\hShift{\hSector q}
\global\long\def\hShiftVec{\hSector q}
\global\long\def\hCC{\hSector{\bar{\mathcal{C}}}}

\global\long\def\sLs{\lSector{\hat{\Lambda}}}
\global\long\def\sLt{\lSector{\lambda}}
\global\long\def\sLtinv{\lSector{\lambda^{-1}}}
\global\long\def\sLv{\lSector v}
\global\long\def\sLp{\lSector p}
\global\long\def\sRs{\lSector{\hat{R}}}
\global\long\def\sRbar{\lSector{\bar{R}}}

\global\long\def\sI{\lSector{\hat{I}}}
\global\long\def\sEta{\lSector{\hat{\delta}}}

\global\long\def\betap#1{\beta_{{\scriptscriptstyle (#1)}}}
\global\long\def\betaScale{\ell^{-2}}
\global\long\def\betaSum{\betaScale{\textstyle \sum_{n}}\beta_{(n)}}
\global\long\def\betaSumL{\betaScale{\displaystyle \sum_{n=0}^{4}}\beta_{(n)}}

\global\long\def\signV{\,+\,}
\global\long\def\isignV{\,-\,}
\global\long\def\usignV{}
\global\long\def\uisignV{-\,}

\global\long\def\isignV{\,+\,}
\global\long\def\signV{\,-\,}
\global\long\def\uisignV{}
\global\long\def\usignV{-\,}

\global\long\def\signK{\,+\,}
\global\long\def\isignK{\,-\,}
\global\long\def\usignK{}
\global\long\def\uisignK{-\,}

\global\long\def\isignK{\,+\,}
\global\long\def\signK{\,-\,}
\global\long\def\uisignK{}
\global\long\def\usignK{-\,}

\global\long\def\signKV{\,+\,}
\global\long\def\isignKV{\,-\,}
\global\long\def\usignKV{}
\global\long\def\uisignKV{-\,}

\global\long\def\isignKV{\,+\,}
\global\long\def\signKV{\,-\,}
\global\long\def\uisignKV{}
\global\long\def\usignKV{-\,}

\global\long\def\signKV{\,+\,}
\global\long\def\isignKV{\,-\,}
\global\long\def\usignKV{+}
\global\long\def\uisignKV{-\,}

\global\long\def\isignKV{\,+\,}
\global\long\def\signKV{\,-\,}
\global\long\def\uisignKV{}
\global\long\def\usignKV{-\,}

\global\long\def\signKV{\,+\,}
\global\long\def\isignKV{\,-\,}
\global\long\def\usignKV{}
\global\long\def\uisignKV{-\,}

\global\long\def\hrD{\hrColor D}
\global\long\def\hrQ{\hrColor Q}
\global\long\def\hrn{\hrColor n}
\global\long\def\hrDn{\hrColor{Dn}}
\global\long\def\hrx{\hrColor x}

\global\long\def\hrV{\hrColor V}
\global\long\def\hrU{\hrColor U}
\global\long\def\hrVbar{\hrColor{\bar{V}}}
\global\long\def\hrWbar{\hrColor{\bar{W}}}
\global\long\def\hrSV{\hrColor S}
\global\long\def\hrUtilde{\hrColor{\tilde{U}}}
\global\long\def\hrVubar{\hrColor{\underbar{V}}}

\global\long\def\hrD{\mathsfit{D}{\mkern1mu}}

\global\long\def\CN{\gSector{\mathcal{C}}}
 \global\long\def\CNdot{\dot{\gSector{\mathcal{C}}}}
 \global\long\def\gCvec{\gSector{\mathcal{C}}}
\global\long\def\CNsm#1{\CN[#1]}

\global\long\def\CL{\fSector{\widetilde{\fSector{\mathcal{C}}}}}
 \global\long\def\CLdot{\dot{\CL}}
 \global\long\def\fCvec{\fSector{\widetilde{\fSector{\mathcal{C}}}}}
\global\long\def\CLsm#1{\CL[#1]}

\global\long\def\Ctwo{\mSector{\mathcal{C}_{2}}}
 \global\long\def\Ctwodot{\dot{\mSector{\mathcal{C}}}_{\mSector 2}}
\global\long\def\Ctwosm#1{\Ctwo[#1]}

\global\long\def\gfCvec{\mSector{\mathcal{R}}}

\global\long\def\fsm{\xi}
\global\long\def\ksm{\eta}

\global\long\def\Cbim{\mSector{\mathcal{C}_{\mathrm{b}}}}

\global\long\def\gW{\gSector{W_{g}}}
 \global\long\def\fW{\fSector{W_{f}}}

\global\long\def\gCW{\gSector{\Omega_{g}}}
 \global\long\def\fCW{\fSector{\Omega_{f}}}

\global\long\def\fWA{A}
\global\long\def\fWB{B}
\global\long\def\fWC{C}
\global\long\def\fWD{D}

\fi


\newcommand{\myTitle}{Note on bimetric causal diagrams}

\newcommand{\myAbstract}{A new kind of diagrams is presented, showing
the causal structure of bimetric interactions.}

\title{\myTitle}

\author{Mikica Kocic}

\affiliation{%
  Department of Physics \& The Oskar Klein Centre,\\
  Stockholm University, AlbaNova University Centre,
  SE-106 91 Stockholm
}

\email{mikica.kocic@fysik.su.se}

\hypersetup{
  pdftitle=\myTitle,
  pdfauthor=Mikica Kocic,
  pdfsubject=Hassan-Rosen ghost-free bimetric theory,
  pdfkeywords={Modified gravity, Massive gravity, 
    Ghost-free bimetric theory, Causal structure}
}

\notoctrue
\emitFrontMatter


\paragraph*{Introduction.}

\begin{figure}[H]
\noindent \begin{centering}
\begin{tikzpicture}[x=1mm,y=1mm]
\node[anchor=south west, inner sep=0] at (0,4) { 
\includegraphics[width=92mm]{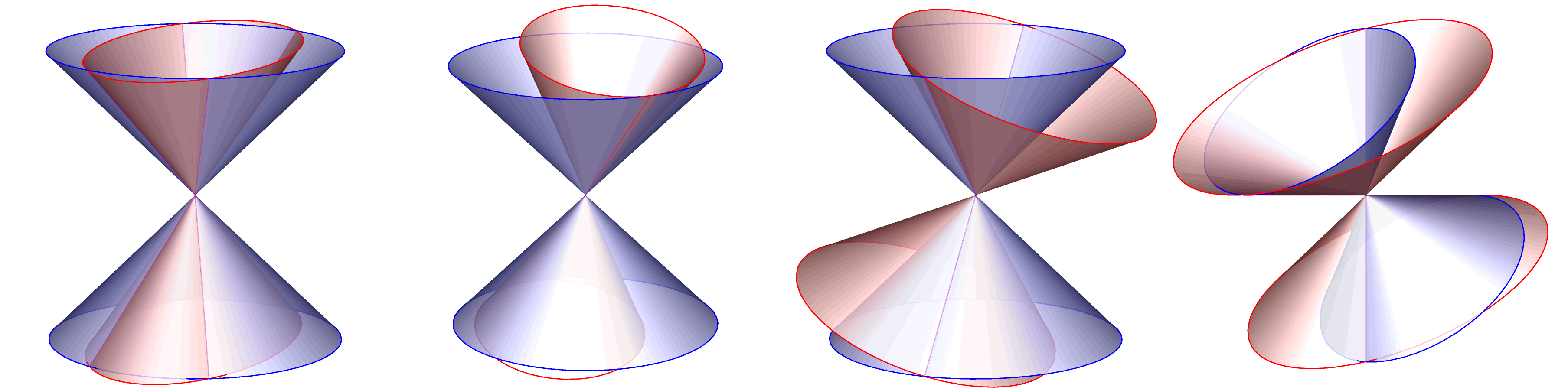}
};
\node[] at (0*23+10,0) {\small Type I};
\node[] at (1*23+10,0) {\small Type IIa};
\node[] at (2*23+10,0) {\small Type IIb};
\node[] at (3*23+10,0) {\small Type III};
\end{tikzpicture}\vspace{-2ex}
\par\end{centering}
\caption{\label{fig1}Null cones for possible metric configurations in the
ghost-free bimetric theory and massive gravity \cite{Hassan:2017ugh}.
Types IIa, IIb, and III are nonbidiagonalizable (the metrics cannot
be simultaneously diagonalized). Type III is absent in spherical symmetry
because of dimensional reduction.}
\end{figure}
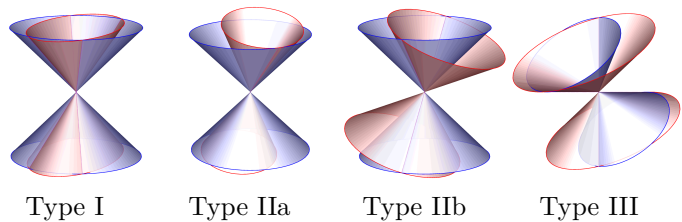

The metric configurations in de Rham\textendash Gabadadze\textendash Tolley
massive gravity \cite{deRham:2010ik,deRham:2010kj,Hassan:2011hr}
and Hassan-Rosen bimetric theory \cite{Hassan:2011zd,Hassan:2018mbl},
can be classified in terms of the intersections of their null cones
\cite{Hassan:2017ugh} (cf.\,\cite{Blas:2005yk}). The classification
of causal types is shown in figure~\ref{fig1}.

\vspace{-1ex}

\begin{figure}[H]
\noindent \begin{centering}
\includegraphics[width=12cm]{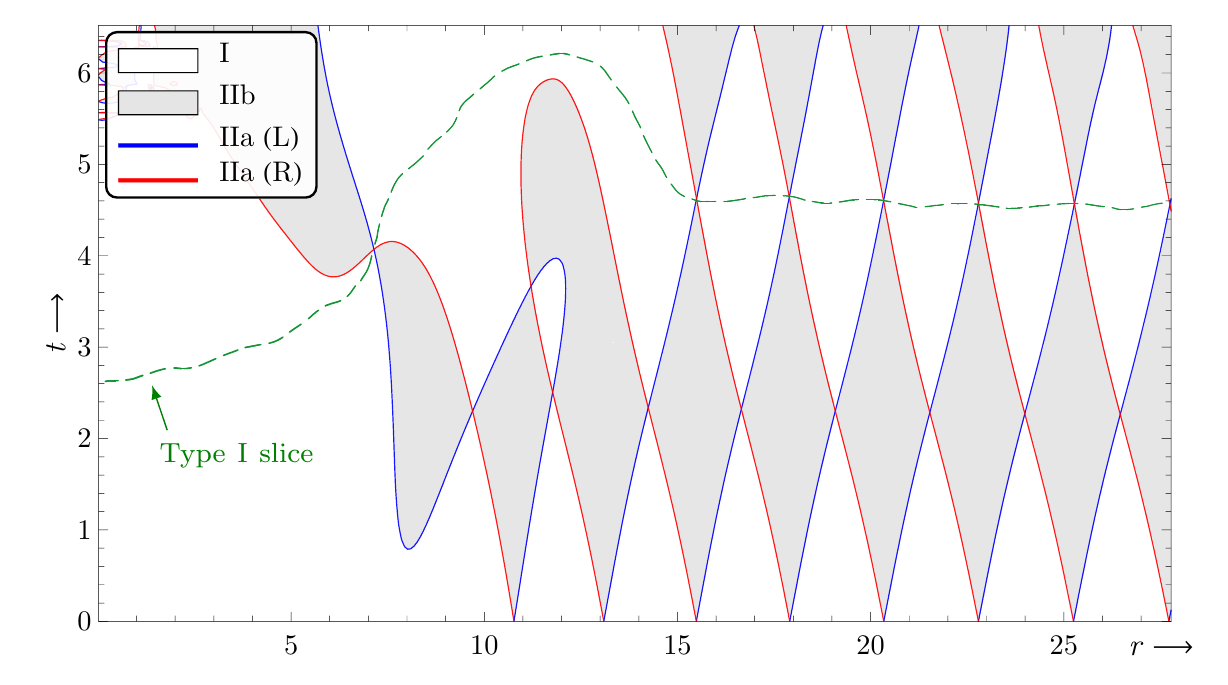}\vspace{-3ex}
\par\end{centering}
\caption{\label{fig2}Causal diagram for a bimetric solution from \cite{Kocic:2019gxl}.
The white regions are Type I, the shaded Type IIb, and the edges Type
IIa. The dashed line is a bidiagonal (Type I) slice.}
\end{figure}

An arbitrary coordinate transformation can only deform the null cones,
keeping the nature of their intersections. Therefore, each point of
bimetric spacetime carries a definite causal type which can be plotted
in a diagram (as an indicator function), displaying an invariant picture
of bimetric spacetime. An example is shown in figure~\ref{fig2},
which illustrates a bimetric spherical dust collapse from~\cite{Kocic:2019gxl}.
The patches in the causal diagram highlight the relative dynamics
of the metric fields showing their oscillations in space and time.

\paragraph*{Construction.}

This section gives a procedure on how to plot a bimetric causal diagram
in spherical symmetry, where every point in the diagram corresponds
to a two-dimensional sphere. Consider two metrics of the following
form,
\begin{align*}
\gMet & =-\gAlpha^{2}\dd t^{2}+\gEA^{2}(\dd r+\gBeta\,\dd t)^{2}+\gEB^{2}\dd\Omega^{2},\\
\fMet & =-\fAlpha^{2}\dd t^{2}+\fEA^{2}(\dd r+\fBeta\,\dd t)^{2}+\fEB^{2}\dd\Omega^{2}.
\end{align*}
Let $\gSector{r_{L}}$ and $\gSector{r_{R}}$ be the coordinates of
the $\gMet$ null cone, and similarly $\fSector{\widetilde{r}_{L}}$
and $\fSector{\widetilde{r}_{R}}$ the coordinates of the $\fMet$
null cone, both cut at $\partial/\partial t$ in the tangent space
(these coordinates can be obtained by setting $\gMet,\fMet,\dd\Omega=0$,
$\dd t=1$, and then solving for $\dd r$),
\begin{alignat*}{3}
\gSector{r_{L}} & \coloneqq-\gBeta-\gAlpha/\gEA, & \gSector{r_{R}} & \coloneqq-\gBeta+\gAlpha/\gEA, & \gSector{r_{L}} & <\gSector{r_{R},}\\
\fSector{\widetilde{r}_{L}} & \coloneqq-\fBeta-\fAlpha/\fEA, & \qquad\fSector{\widetilde{r}_{R}} & \coloneqq-\fBeta+\fAlpha/\fEA, & \qquad\fSector{\widetilde{r}_{L}} & <\fSector{\widetilde{r}_{R}}.
\end{alignat*}
Then, the causal type can be determined from the following two distances
using table \ref{tab1},
\[
L_{\Delta}\coloneqq\gSector{r_{L}}-\fSector{\widetilde{r}_{L}},\quad R_{\Delta}\coloneqq\gSector{r_{R}}-\fSector{\widetilde{r}_{R}}.\vspace{-5mm}
\]

\begin{table}[H]
\noindent \centering{}\caption{\label{tab1}Causal type discrimination based on $L_{\Delta}$ and
$R_{\Delta}$.}
\vspace{1ex}{\begingroup\renewcommand{\arraystretch}{1.15}\small%
\begin{tabular}{cccll}
\hline 
$L_{\Delta}$ & $R_{\Delta}$ & $L_{\Delta}R_{\Delta}$ & Causal type & Comment\tabularnewline
\hline 
\hline 
$<0$ & $<0$ & $>0$ & Type IIb & $\gMet$ left of $\fMet$\tabularnewline
$<0$ & $=0$ & $=0$ & Type IIa (R) & right null direction in common\tabularnewline
$<0$ & $>0$ & $<0$ & Type I & $\fMet$ inside $\gMet$ \tabularnewline
$=0$ & $<0$ & $=0$ & Type IIa (L) & left null direction in common\tabularnewline
$=0$ & $=0$ & $=0$ & Type I & double null direction in common\tabularnewline
$=0$ & $>0$ & $=0$ & Type IIa (L) & left null direction in common\tabularnewline
$>0$ & $<0$ & $<0$ & Type I & $\gMet$ inside $\fMet$\tabularnewline
$>0$ & $=0$ & $=0$ & Type IIa (R) & right null direction in common\tabularnewline
$>0$ & $>0$ & $>0$ & Type IIb & $\fMet$ left of $\gMet$\tabularnewline
\hline 
\end{tabular}\endgroup}
\end{table}

\begin{figure}[H]
\noindent \begin{centering}
\includegraphics{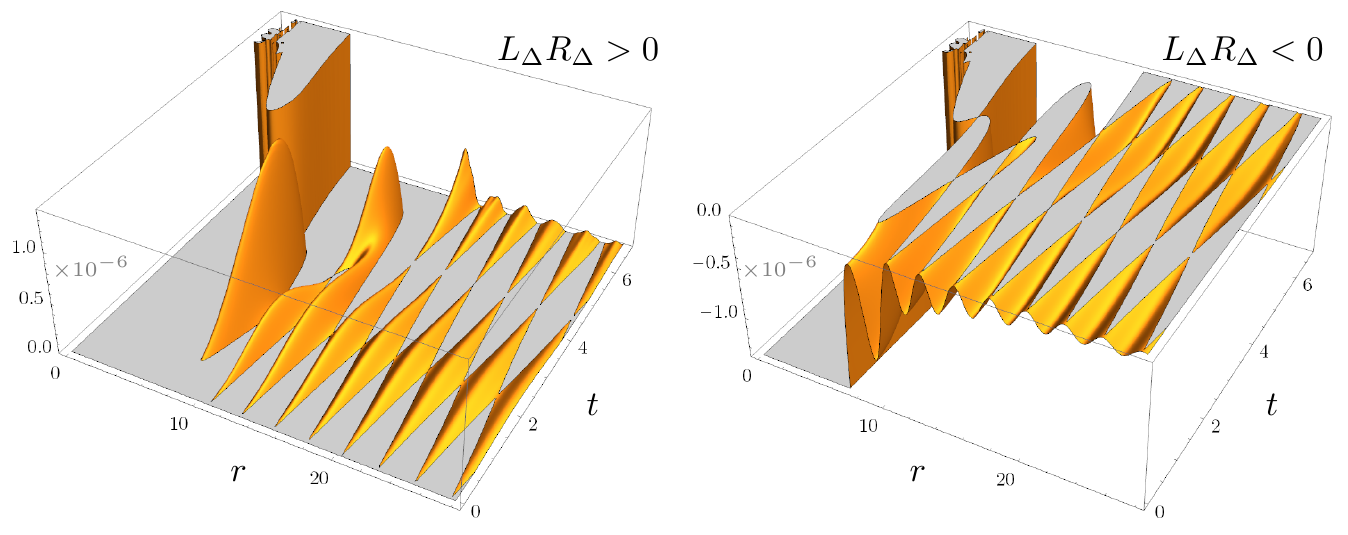}\vspace{-3ex}
\par\end{centering}
\caption{\label{fig3}Discriminant $L_{\Delta}R_{\Delta}$ for the bimetric
solution in figure~\ref{fig2}.}
\end{figure}

\noindent The causal regions can be shown by plotting the product
$L_{\Delta}R_{\Delta}$, in which case Type I regions have $L_{\Delta}R_{\Delta}<0$,
Type IIb regions $L_{\Delta}R_{\Delta}>0$, and Type IIa regions have
$L_{\Delta}R_{\Delta}=0$ (Type I if it is a double zero). An example
for the solution from figure~\ref{fig2} is shown in figure~\ref{fig3}.

\paragraph*{Outlook.}

\begin{figure}[H]
\noindent \begin{centering}
\includegraphics[scale=0.9]{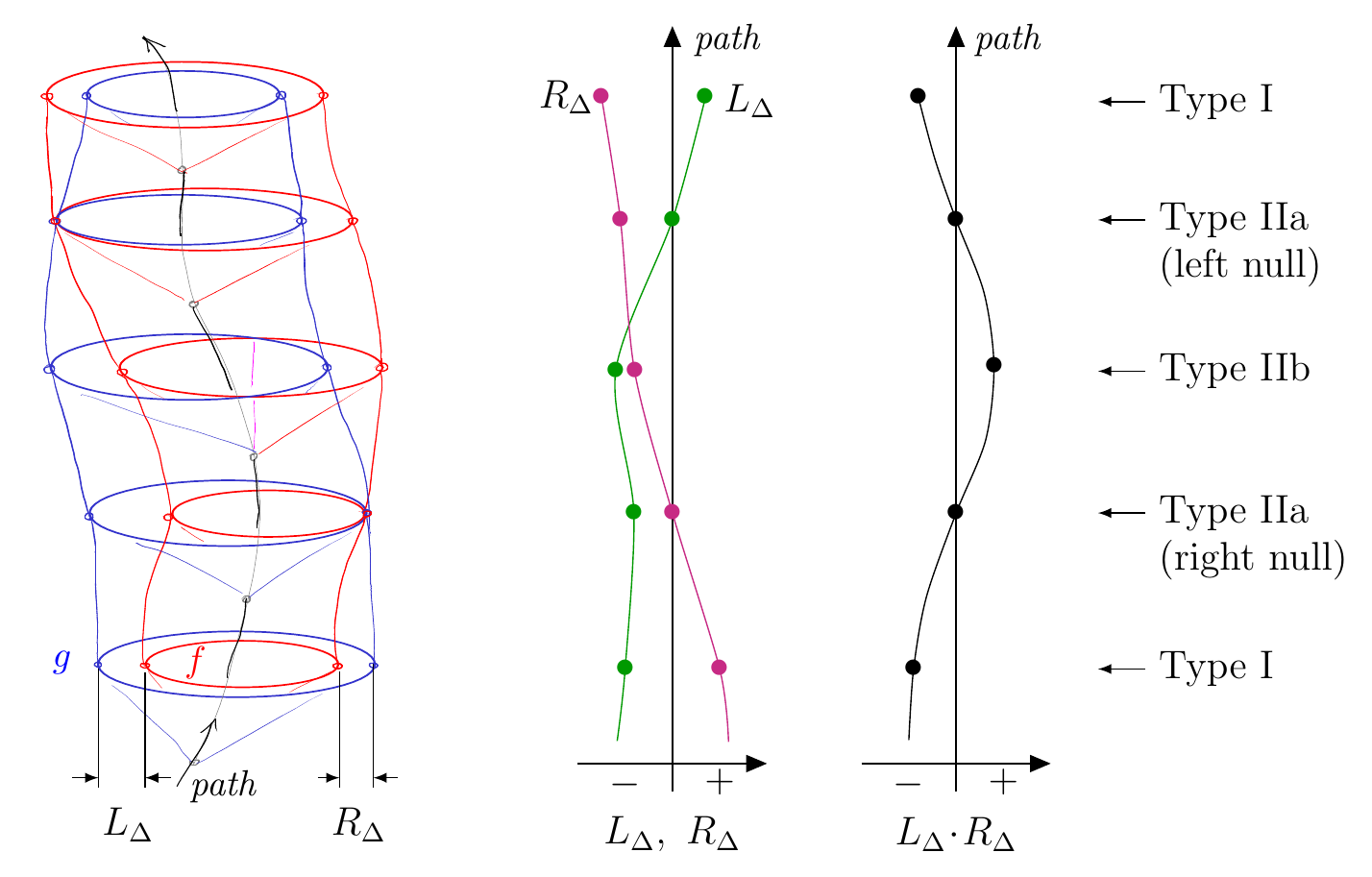}\vspace{-3ex}
\par\end{centering}
\caption{\label{fig4}Tracking the causal type along an arbitrary path.}
\end{figure}

Note that $L_{\Delta}$ and $R_{\Delta}$ are smooth functions. Besides,
the coordinate transformations can only smoothly deform the causal
diagram: the imprinted topological features, such as which regions
are adjacent to each other, can not be removed. Consequently, one
can construct a ``painted'' Penrose\textendash Carter diagram for
each sector, depicting the relative causal structure between the two
metrics. Finally, the causal types can be tracked along any curve
(or surface) as shown in figure~\ref{fig4}. For example, bidiagonal
slices comprise Type~I paths (not necessarily spacelike), illustrated
by the dashed line in figure~\ref{fig2}.\vspace{-1ex}

\vspace{1ex}

\noindent \textbf{Acknowledgments.} I wish to thank Fawad Hassan,
Edvard M\"{o}rtsell, Francesco Torsello, Anders Lundkvist, and Marcus
H\"{o}g\r{a}s for the discussions and helpful comments.

\section*{References}\vspace{-1ex}
\begingroup\renewcommand{\section}[2]{}%

\bibliographystyle{JHEP}
\bibliography{bim-cd}

\providecommand{\href}[2]{#2}\begingroup\raggedright\begin{thebibliography}{1}

\bibitem{deRham:2010ik}
C.~de~Rham and G.~Gabadadze, \emph{{Generalization of the Fierz-Pauli Action}},
  \href{https://doi.org/10.1103/PhysRevD.82.044020}{\emph{Phys. Rev.}
  {\bfseries D82} (2010) 044020},
  [\href{https://arxiv.org/abs/1007.0443}{{\ttfamily 1007.0443}}].

\bibitem{deRham:2010kj}
C.~de~Rham, G.~Gabadadze and A.~J. Tolley, \emph{{Resummation of Massive
  Gravity}}, \href{https://doi.org/10.1103/PhysRevLett.106.231101}{\emph{Phys.
  Rev. Lett.} {\bfseries 106} (2011) 231101},
  [\href{https://arxiv.org/abs/1011.1232}{{\ttfamily 1011.1232}}].

\bibitem{Hassan:2011hr}
S.~F. Hassan and R.~A. Rosen, \emph{{Resolving the Ghost Problem in non-Linear
  Massive Gravity}},
  \href{https://doi.org/10.1103/PhysRevLett.108.041101}{\emph{Phys. Rev. Lett.}
  {\bfseries 108} (2012) 041101},
  [\href{https://arxiv.org/abs/1106.3344}{{\ttfamily 1106.3344}}].

\bibitem{Hassan:2011zd}
S.~F. Hassan and R.~A. Rosen, \emph{{Bimetric Gravity from Ghost-free Massive
  Gravity}}, \href{https://doi.org/10.1007/JHEP02(2012)126}{\emph{JHEP}
  {\bfseries 02} (2012) 126},
  [\href{https://arxiv.org/abs/1109.3515}{{\ttfamily 1109.3515}}].

\bibitem{Hassan:2018mbl}
S.~F. Hassan and A.~Lundkvist, \emph{{Analysis of constraints and their algebra
  in bimetric theory}},
  \href{https://doi.org/10.1007/JHEP08(2018)182}{\emph{JHEP} {\bfseries 08}
  (2018) 182}, [\href{https://arxiv.org/abs/1802.07267}{{\ttfamily
  1802.07267}}].

\bibitem{Hassan:2017ugh}
S.~F. Hassan and M.~Kocic, \emph{{On the local structure of spacetime in
  ghost-free bimetric theory and massive gravity}},
  \href{https://doi.org/10.1007/JHEP05(2018)099}{\emph{JHEP} {\bfseries 05}
  (2018) 099}, [\href{https://arxiv.org/abs/1706.07806}{{\ttfamily
  1706.07806}}].

\bibitem{Blas:2005yk}
D.~Blas, C.~Deffayet and J.~Garriga, \emph{{Global structure of bigravity
  solutions}}, \href{https://doi.org/10.1088/0264-9381/23/5/015}{\emph{Class.
  Quant. Grav.} {\bfseries 23} (2006) 1697--1719},
  [\href{https://arxiv.org/abs/hep-th/0508163}{{\ttfamily hep-th/0508163}}].

\bibitem{Kocic:2019gxl}
M.~Kocic, F.~Torsello, M.~H\"{o}g\r{a}s and E.~Mortsell, \emph{{Spherical dust
  collapse in bimetric relativity: Bimetric polytropes}},
  \href{https://arxiv.org/abs/1904.08617}{{\ttfamily 1904.08617}}.

\end{thebibliography}\endgroup

\endgroup

\end{document}